\documentclass[a4paper,11pt]{article}
\pdfoutput=1 

\usepackage{jcappub} \usepackage{graphicx}	
\usepackage{amsmath}	
\usepackage{amssymb}	
\usepackage{commath}
\usepackage{ulem}
\usepackage{comment}
\usepackage{multirow}
\usepackage{bigints}
\usepackage{etoolbox}
\usepackage{mathtools}
\usepackage[T1]{fontenc} 
\usepackage{orcidlink}

\title{\boldmath Dynamical dark energy models in the light of Gravitational-Wave Transient Catalogues}

\author[a,1]{Celia Escamilla-Rivera\orcidlink{0000-0002-8929-250X}}
\author[a]{and Antonio N\'ajera\orcidlink{0000-0001-9738-7704}}

\affiliation[a]{Instituto de Ciencias Nucleares, Universidad Nacional
	Aut\'onoma de M\'exico, Circuito Exterior C.U., A.P. 70-543,
	M\'exico D.F. 04510, M\'exico}

\emailAdd{celia.escamilla@nucleares.unam.mx} 
\emailAdd{antonio.najera@correo.nucleares.unam.mx}

\abstract{The study of current gravitational waves (GW) catalogues provide an interesting model independent way to understand further the \textit{nature} of dark energy. In this work, we present an update of the constrains related to dynamical dark energy parametrisations using recent Gravitational-Wave Transient catalogues (GWTC-1 and GWTC-2) along with Type Ia supernova (SNeIa) and Cosmic Chronometers (CC) catalogues. According to our Bayesian results using the full SNeIa+CC+GW database, the  $\Lambda$CDM model shows a strong preference against two dark energy parameterisation known as Barboza-Alcaniz (BA) and the Low Correlation (LC) models. Also, we obtain a very strong preference against the Chevallier-Polarski-Linder (CPL) model. Furthermore, we generated a mock GW catalogue and estimate that we require approximately 1000 standard sirens to have a constrain of $H_0$ within 1\% relative error, quantity that is out of reach of current standard sirens candidates in GWTC-1 and GWTC-2 catalogues.}

\begin{document}
	\keywords{Dark energy -- gravitational waves -- precision cosmology}
	
\maketitle
\flushbottom


\section{Introduction}
\label{sec:intro}
One of the current challenges of precision cosmology is to understand the nature behind the late-cosmic acceleration. The study of a wide variety of observations, e.g. Cosmic Microwave Background Radiation (CMB), Supernovas Type Ia (SNeIa), Baryonic Acoustic Oscillations (BAO) and the current Gravitational Waves (GWs) have been useful to constrain the cosmological parameters that are defined in a specific cosmological model.  Among these observations, some of them are consistent with the standard $\Lambda$CDM model, which has its own theoretical problems, e.g. the Hubble constant tension issue \cite{DiValentino:2020zio}, and references therein, or the cosmological constant problem related with the disagreement between the observed values of vacuum energy density and theoretical large value of zero-point energy suggested by quantum field theory. Therefore, it is worthwhile to consider alternative options. 
A convenient road is to keep General Relativity (GR) as the gravitational theory and modify the stress-energy content of the universe, i.e., we modify the right-hand side in Einstein equations. A second road is to design an alternative theory of gravity, whose additional degrees of freedom can handle the cosmic acceleration --so-called the modified gravity method-- via modifying the left-hand side of the Einstein equations, which leads to new physics on small cosmological scales.

On the other hand, one straightforward way to explore a specific cosmic accelerated expansion is at phenomenological level. In such case, we can quantify the evolution of an equation of state (EoS) parametrisation (geometric in the case of modified gravity) of dark energy. Given the information on the values of the Hubble parameter $H_0$, and the current matter density fraction, $\Omega_m$, an arbitrary expression for the expansion evolution can be reproduced by assuming a flat Friedmann-Lemaître-Robertson-Walker (FLRW) space time with a dark energy component that has a characteristic EoS $w(z)$. Scenarios were a possible $w(z)\neq 1$ have been discussed extensively and this denotes a strong evidence of new gravitational physics and a possible new source of the cosmic acceleration. Fitting a constant $w$ using observational data results in good agreement with an EoS of $w=-1$, but such analysis would have missed subtle variations in the EoS, e.g. the average value near to $w=-1$. Likewise, using a specific parametric form of $w(z)$ is liable to produce a bias on the outcome. In this line of thought, some proposals have been currently studied, e.g. bidimensional dark energy models \cite{Escamilla-Rivera:2016qwv}, EoS evolving with a scalar field dark energy \cite{Escamilla-Rivera:2016aca}, generic EoS derived from $f(R)$ \cite{Jaime:2018ftn,Escamilla-Rivera:2020giy,Cruz:2020cje}, double exponential potentials for late-cosmic acceleration \cite{Das:2018xvj}, interacting dark energy models like $w(q)$ \cite{Elizalde:2018ahd}, Pad\'e approximants for dark energy \cite{Rezaei:2017yyj}, equivalent description of dark energy as cosmographic test \cite{Bamba:2012cp}
and dark energy inverse cosmography \cite{Escamilla-Rivera:2019aol,Munoz:2020gok}. 

In the next decade, future proposals will need to be addressed  by taking into account the current discordance between the different cosmological probes \cite{DiValentino:2020vhf}. Recently, constraints derived from considering the interplay between GWs and other light gravitational degrees of freedom, related to dark energy, have been used to strongly restrict the latter \cite{Mukherjee:2020mha,Mukherjee:2020hyn}. Also, using the electromagnetic transient characteristic from the first multi-messenger observations of a binary neutron star merger \cite{GBM:2017lvd,Mukherjee:2020kki},  a possibility to constraint modified gravities have emerged. In this matter, several classes of scalar-tensor theories were found incompatible with the GW data or else irrelevant to gravity at sufficiently large distances \cite{Ezquiaga:2018btd,Baker:2020apq}. Some loopholes have been discussed by scanning a wide range of modified gravity theories such as Beyond Horndeski and DHOST \cite{Bordin:2020fww}. 

The goal of this work is to present an update in the study of dark energy parametrisations by focusing on the new GW data, using Gravitational-Wave Transient Catalogues of Compact Binary Mergers, observed by LIGO and Virgo during the first, second, and the first half of the third observing runs \cite{abbott2019gwtc,Abbott:2020niy} in order to shed some light on the cosmological constraints. In this line of thought, we require a model-independent determination of the redshift $z$ up to the GW source.
This determination either requires an electromagnetic (EM) counterpart of the GW source, the localisation of the GW source galaxy or another method to assign it in a specific redshift $z$ range. Apart from using current standard sirens (gravitational wave sources with an independent measurement of their redshift), in this paper we use an alternative version, which was proposed by Ding et al.  in \cite{ding2019cosmological}, where instead of calculating $z$, a prior redshift probability is used. The posterior of cosmological parameters with GW data can be written using the Bayes theorem through the total probability rule $\mathcal{P}$ in terms of the likelihood $\mathcal{L}$, the prior probability of the parameters, and the redshift prior probability. The redshift prior will be taken as the compact binary merger rate, which allows us to perform precision analyses without the need of EM counterparts for the sources. These GW events without EM counterparts are called dark sirens. In addition to this task, we will simulate standard sirens to see how many of those events we need to constrain the Hubble constant $H_0$ with a relative error within 1\%, taking into account  the standard $\Lambda$CDM model.

This paper is organised as follows: 
	In Sec.\ref{sec:sirenDistance}, we describe general properties of the standard sirens (the distance to GW sources).
	In Sec.\ref{sec:DEbackground}, it is present the theory behind dark energy parametrisations. Some alternative bidimensional parametrisations are considered for the EoS, which deviate from the standard $\Lambda$CDM $w(z) = -1$. Among them, we will study the Chevallier-Polarski-Linder (CPL), the Barboza-Alcaniz (BA) and the Low Correlation (LC) parametrisations. In Sec.\ref{sec:cosmologyDarkSirens}, we  discuss the Bayesian method to be performed for the case of dark sirens.
	In Sec.\ref{sec:data}, the observational compilations considered in this work are described. Furthermore, we will explain the generation of the mock standard sirens according to these catalogues.
	In Sec.\ref{sec:results}, we will perform the statistical analysis on the dark energy models discussed using the observational compilations including the GW catalogues.
	Finally, in Sec.\ref{sec:conclusions} we present our main results.

\section{Standard Sirens}
\label{sec:sirenDistance}

Since GWs have energy and hence flux and luminosity, the definition of this distance is analogous to the luminosity distance. Due to this, we will introduce this definition before dealing with the standard siren distant and also because, as we will see, the latter is written in terms of the former one.

\subsection{Luminosity distance}
\label{sec:background}

To compute the luminosity distance to a source, it is required to measure its flux $F$ for an object of known luminosity $L$. The luminosity distance $D_L$ is
\begin{equation}
\label{eqn:luminosityDistance}
D_L = \left(\frac{L}{4\pi F}\right)^{1/2},
\end{equation}
which can be written assuming a spatially flat universe as \cite{dodelson2020modern}
\begin{equation}
\label{eqn:luminosityDistanceComoving}
D_L(z) = (1+z) \, D_H \int_{0}^z \dfrac{dz'}{E(z')},
\end{equation}
where $D_H = c/H_0$ is the Hubble distance ($c$ is the speed of light and $H_0$ the Hubble constant) and $E(z) = H(z)/H_0$ (with $H(z)$ the Hubble parameter). This distance enables us to test cosmological models with sources of electromagnetic radiation. Because of that, to test those models with GWs, we need a theoretical expression for the distance to a GW source analogous to (\ref{eqn:luminosityDistanceComoving}).  \\

\subsection{Theoretical standard sirens}

While the employ of distances in Cosmology are usually defined using the light detected from distant sources, GWs give us new useful tools for measuring
cosmological distances. In that regards, we can consider: 
\begin{itemize}
	\item The {\it GW luminosity distance ($d_L^{\rm GW}$)}. 
	Defined in terms of the ratio of GW power emitted at a source position versus the GW flux at the detector.
	The luminosity distance depends on the cosmic expansion rate and can be directly measured from a detection of a  GW from distant sources. This quantity results of interest since it can be used to test theories of modified gravity \cite{Ezquiaga:2018btd}.
	\item { The {\it GW angular distance ($d_A^{\rm (GW)}$)}. Defined in terms of the ratio between the source angular diameter at emission versus the source angular size at detector. Unfortunately, the angular resolution is not well precise in current detectors. Moreover, we have already knowledge about the progress in this line of thought \cite{Baker:2019ync}.}
\end{itemize}
Therefore, to found an explicit relation between both distances, $d_A$ can be related with $d_L$ through the duality-distance relation $d_L\,=\,(1+z)^2\,d_A$. Moreover, the GW evolution can be affected by a friction term which depends on slowly-varying fields only when we take into account a linearised energy-momentum tensor. At this point, we can expect that both distances are influenced by modified gravity and these quantities are related  between them
by the duality-distance relation rewritten as
\begin{equation}
d_L^{\rm GW}\,=\,(1+z)^2\,d_A^{\rm GW}\,,
\end{equation}
where the GW is propagating through a perturbed FLRW space time in scalar-tensor scenarios. This $d_L^{\rm GW}$ is what we call \textit{standard siren distance} $D_S$. A possible form of the relationship between the standard luminosity and the standard siren distance is given by
\cite{Mitra:2020vzq}
\begin{equation}
\label{eqn:sirenDistance}
D_S(z) = D_L(z) \exp \left( - \int_{0}^{z} \frac{\delta(z')}{1+z'} dz' \right),
\end{equation}
which comes from modified gravity models  \cite{belgacem2019testing}. The subscript $S$ and $L$ denotes the siren luminosity distance and the standard candle luminosity distance, respectively. In this work, we will deal with dynamical dark energy models from GR. Hence, we will consider $\delta(z) = 0$. In this framework, the siren and luminosity distances are equivalent.

\section{Dynamical dark energy parametrisations}
\label{sec:DEbackground}

Since the concordance cosmological $\Lambda$CDM model has several theoretical problems \cite{Weinberg:2000yb}, new proposals have been presented along these years. One of the most popular is related to the addition of an exotic dynamic fluid denoted as dynamical dark energy. 
Several dark energy EoS $w$ have been built to describe the late cosmic acceleration. In order to see which one of them has more constraining power, we test them using observational data. 

The fact that the Universe is experimenting an accelerated expansion directs us to consider a negative pressure component at late times. Since energy is always positive, the dark energy EoS $w=w(z)$ requires to be $w(z)=P/\rho<-1/3$. This can be seen from the second Friedmann equation, $\ddot{a}/a = -4\pi G/3 (1+3w)\rho$. Then, to have an accelerated cosmic expansion, we require $w(z)<-1/3$. For a spatially flat universe 
\begin{equation}
\label{eqn:FriedmannEquation}
E(z)^2=\left(\frac{H(z)}{H_0}\right)^2 = \Omega_m (1+z)^3 + \Omega_\Lambda \, f(z),
\end{equation}
where the function $f(z)$ is defined as
\begin{equation}
\label{eqn:darkEnergyAnsatz}
f(z) = \exp \left( 3 \int_0^z \frac{1+w(z')}{1+z'} dz' \right).
\end{equation}

Therefore, for a given $w(z)$, the dynamics of the expansion are determined. The $\Lambda$CDM model is a special case when $w(z)=-1$ and $f(z)=1$. 
From here, we can consider the following three bidimensional\footnote{We denoted \textit{bidimensional} as the dependence solely of two free theoretical parameters $(w_0,w_1)$.} dark energy parametrisations \footnote{A previous analysis of these models has been presented in \cite{Escamilla-Rivera:2016qwv}.} with different analytical forms of $w(z)$.

\subsection{Lambda Cold Dark Matter ($\Lambda$CDM) model}

This is the standard model of Cosmology denoted by a EoS with $w(z)=-1$. The Friedmann equation for this model becomes
\begin{equation}
E(z)^2=\Omega_m (1+z)^3+(1-\Omega_m).
\end{equation}
This is a simple model that describes the late-time cosmic acceleration.


\subsection{Chevallier-Polarski-Linder (CPL) case}
\label{subsec:CPLParametrization}

This parametrisation considers an EoS \cite{Chevallier:2000qy,Linder:2007wa} of the form
\begin{equation}
\label{eqn:CPLEOS}
w(z) = w_0 + \left(\frac{z}{1+z}\right) w_1,
\end{equation}
which converges for high redshifts and reduces to the $\Lambda$CDM model for  $w_0=-1$ and $w_1=0$. For this case, the Friedmann equation is given by
\begin{eqnarray}
\label{eqn:FriedmannEquationCPLParametrization}
E(z)^2 = \Omega_m (1+z)^3 + (1-\Omega_m) \exp\left( -\frac{3 w_1 z}{1+z} \right) (1+z)^{3(1+w_0+w_1)}.
\end{eqnarray}


\subsection{Barboza-Alcaniz  (BA) case}

This model proposes an EoS \cite{barboza2008parametric} of the form
\begin{equation}
w(z)=w_0 + \frac{z(1+z)}{1+z^2} w_1.
\end{equation}
Notice that this expression does not diverge for $z\to -1$. This re-scaling is helpful when considering the entire history of the universe because of the definition of $a={1}/{(1+z)}$, with $a \in [0,\infty)$ and hence $z \in (-1,\infty)$. In this case, the Friedmann equation is 
\begin{eqnarray}
E(z)^2 = \Omega_m (1+z)^3 + (1-\Omega_m) (1+z)^{3(1+w_0)} (1+z^2)^{3w_1/2}.
\end{eqnarray}


\subsection{Low Correlation  (LC) case}

This is a model with an EoS \cite{wang2008figure} given by
\begin{equation}
w(z) = \frac{(-z+z_c)w_0+z(1+z_c)w_c}{(1+z)z_c},
\end{equation}
where $w_0=w(0)$ and $w_c=w(z_c)$ and $z_c$ is a given redshift value so that $(w_0,w_z)$ are uncorrelated. These values depend on the data used to be constrained. In this work, we considered $z_c = 0.5$, which is a conservative value of this parameter that gives low correlation between $w_0$ and $w_c$.
This parametrisation reduces to $\Lambda$CDM for $w_0 = -1$ and $w_c = -1$ .The Friedmann equation is 
\begin{eqnarray}
E(z)^2 = \Omega_m (1+z)^3 + (1-\Omega_m)
 \exp \left( \frac{9(w_0-w_c)z}{1+z} \right) 
(1+z)^{3(1-2w_0+3w_c)}.
\end{eqnarray}


\section{Cosmology with Dark Sirens}
\label{sec:cosmologyDarkSirens}

In this section, we will use a method to compute the posterior probability of dark sirens proposed by Ding et al. \cite{ding2019cosmological}. Since dark sirens do not have an independent measurement of their redshift, we need to consider an alternative test for our models with these data. We need to compute the posterior probability of the vector of parameters $\Theta$ with the measured siren distances $D_S$, along with their uncertainties $\sigma_{D_S}$, i.e. we want to compute $\mathcal{P}(\Theta | D_S, \sigma_{D_S}, \eta)$, where $\eta$ denotes some astrophysical parameters. First, we consider a single measurement of a siren distance $D_{S_i}$. Then, by applying the Bayes theorem
\begin{equation}
	\label{eqn:BayesTheorem}
	\mathcal{P}(\Theta | D_{S_i}, \sigma_{D_{S_i}},\eta) = \frac{\mathcal{P}(D_{S_i} | \sigma_{D_{S_i}}, \Theta, \eta) \, \mathcal{P}(\Theta)}{\mathcal{P}(D_{S_i} | \sigma_{D_{S_i}},\eta)},
\end{equation}
where $\Theta$ denotes a vector of cosmological parameters $\Theta = \{ h, \Omega_M, w_0, w_1 \}$. We can rewrite $\mathcal{P}(D_{S_i}| \sigma_{D_{S_i}}, \Theta, \eta)$ using the total probability rule, obtaining the following expression
\begin{equation}
	\label{eqn:totalProbability}
	\mathcal{P}(D_{S_i} | \sigma_{D_{S_i}} , \Theta, \eta) = \int_0^\infty dz ~\, \mathcal{P}(D_{S_i} | \sigma_{D_{S_i}} , \Theta, z) \, \mathcal{P}(z | \Theta, \eta),
\end{equation}
where $\mathcal{P}(z|\Theta,\eta)$ is the conditional prior redshift probability on the cosmological/astrophysical parameters. Notice that this prior will depend on the model. 
Inside the integrand of the previous equation, we have the likelihood \cite{hogg2010data}
\begin{eqnarray}
	\label{eqn:likelihood}
	\mathcal{P}(D_{S_i} | \sigma_{D_{S_i}} , \Theta, z) = \frac{1}{\sqrt{2 \pi \sigma_{D_{S_i}}^2}} \exp \left( - \frac{1}{2} \left[ \dfrac{D_S(z, \Theta)-D_{{S_i}. \, \text{obs}}}{\sigma_{D_{S_{i}}}} \right]^2 \right).
\end{eqnarray}
If we have $N$ measured events, the posterior of the full catalogue needs to be consider. All measurements should be independent of each other and hence the posterior of the full dataset is the product of the individual posteriors. Hence, the final posterior probability is given by
\begin{eqnarray}
	\label{eqn:posteriorProbability}
	\mathcal{P}(\Theta|D_s, \sigma_{D_S}) = \frac{\mathcal{P}(\Theta)}{\mathcal{E}} \prod_{n=1}^N \int_0^{z_\text{max}} dz ~\mathcal{P}(D_{S_n}|\sigma_{D_{S_n}}, \Theta, z) ~ \mathcal{P}(z|\Theta,\eta),
\end{eqnarray}
where $\mathcal{E}$ is the evidence, which is a normalisation factor and therefore can be excluded when performing an MCMC.

Notice that we will not expect to detect GW events to a large arbitrary $z$. Therefore, we will perform the integral from 0 up to $z_\text{max}$ being this redshift a reasonable upper limit at which the network detector under consideration can detect GW sources. We will consider $z_{\max}=1.5$ since approximate redshifts of the GWTC-1 \cite{abbott2019gwtc} and GWTC-2 \cite{Abbott:2020niy} catalogues is less than $z=1$ for all their GW events assuming a flat $\Lambda$CDM model. Hence, this value of $z_{\max}=1.5$ is a conservative upper limit. We can compute the likelihood with (\ref{eqn:likelihood}), however, we still need to know the redshift prior $\mathcal{P}(z|\Theta,\eta)$. It can be taken as a probability function both describing the intrinsic merger rate of binaries and the efficiency of the GW network detector under consideration. This is called differential inspiral rate per redshift and it is given by \cite{ding2019cosmological}
\begin{equation}
	\label{eqn:redshiftPrior}
	\frac{d \dot{N}}{dz} = 4 \pi \left(\frac{c}{H_0}\right)^3 \frac{\dot{n}(z)}{1+z} \dfrac{r^2(z)}{E(z)} C_\theta (x(z, \eta)),
\end{equation}
where $\dot{N}$ is the inspiral merger rate, $\dot{n}(z)$ the intrinsic merger rate at redshift $z$, $r$ the comoving adimensional distance and $C_\theta$ a function that measures the network detector performance \cite{biesiada2014strong}. The function $C_\theta$ includes the astrophysical parameters as we will see in the following procedure. \\

Notice that $d\dot{N}/dz$ has units of $\text{yr}^{-1}$, but probability distributions must not have dimensions. Hence, we need to include an additional constant $B$ that will delete the dimensions of the prior. Since the differential inspiral rate per redshift (\ref{eqn:redshiftPrior}) depends on the vector of cosmological parameters $\Theta$ and the vector of astrophysical parameters $\eta$. This latter constant will depend on them as well. Furthermore, probabilities must be normalised. Therefore, $B$ is also a normalisation constant and then
\begin{equation}
	1 = B(\Theta,\eta) \int_{0}^{z_{\text{max}}} dz \dfrac{d\dot{N}}{dz},
\end{equation}
which allows to delete the dimensions of $d\dot{N}/dz$ and to normalise the redshift prior. Hence, we can write the redshift prior probability as
\begin{equation}
	p(z|\Theta,\eta) = B(\Theta,\eta) \frac{d\dot{N}(z,\Theta,\eta)}{dz},
\end{equation}
which is adimensional and properly normalised as all probabilities need to be. The value of the normalisation constant $B(\Theta,\eta)$ must be computed at each iteration of the MCMC to properly compute the redshift prior due to its dependence in the vectors $\Theta$ and $\eta$. When is not consider the normalisation function $B(\Theta,\eta)$ we obtain strong biases on the posterior probabilities of the cosmological parameters. The convergence is not achieved without this constant and the best fits of the parameters tend strongly to the upper or lower limits of the prior uniform probabilities. Moreover, the function $d\dot{N}(z, \Theta,\eta)/dz$ includes the term $C_\theta(x(z,\eta))$, which, as we will see, takes into account the performance of the GW detector. Therefore, this redshift prior probability takes into account the selection effects that affect the inference of the cosmological parameters \cite{mandel2019extracting}.

As we can notice, (\ref{eqn:redshiftPrior}) depends on the parameters $\Theta$, moreover, it also depends on astrophysical parameters $\eta$, which are the signal to noise ratio and the chirp mass in the reference frame of the detector of each GW event. We called the vector of these astrophysical quantities $\eta = \{ \rho, \mathcal{M}_\text{det} \}$. We will consider the intrinsic merger rates $\dot{n}(z)$ computed in \cite{dominik2013double} where they used the population synthesis code. We took the standard low-end model in the observer frame. These results are available as a library of stellar populations\footnote{\url{https://www.syntheticuniverse.org/}}. Finally, the $C_\theta$ function is given by \cite{ding2019cosmological} 
\begin{equation}
	C_\theta(x(z,\eta)) = \int_{x(z, \eta)}^\infty d\theta P_\theta (\theta),
\end{equation}
where
\begin{equation}
	\begin{split}
		P_\theta(\theta) = \begin{cases}
			\dfrac{5}{256} \theta (4-\theta)^3 \qquad \text{if} \quad 0 < \theta < 4, \\ \\
			\qquad 0 \qquad \qquad \qquad \text{otherwise},
		\end{cases} 
	\end{split}
\end{equation}
and
\begin{eqnarray}
	x(z, \rho) = \frac{\rho}{\rho_0} (1+z)\dfrac{c}{H_0} \dfrac{r(z)}{r_0} \left[ \dfrac{1.2 \, M_{\odot}}{\mathcal{M}_\text{det}} \right]^{5/6},
\end{eqnarray}
with $\rho$ the signal to noise measured ratio, $\rho_0$ the signal to noise threshold, $r_0$ the characteristic distance of the network detector and $\mathcal{M}_{\text{det}}$ the chirp mass measured in the reference frame of the detector. Both $\rho$ and $\mathcal{M}_\text{det}$ vary in each GW detection. As it can be seen, with a longer $r_0$, the network detector can detect events at a higher redshift because $x$ should be on the interval $(0,4)$ to have a probability bigger than 0 to detect a given event in a redshift $z$ due the form of the $\mathcal{P}_\theta(\theta)$ function.

In our analyses, we will consider $r_0 = 120 \, \text{Mpc}$ as they are the within the range of the characteristic sensitivities of advanced LIGO for the case of binary black holes (BBH) and $\rho_0 = 8$ \cite{Abbott2020prospects}.


\section{Observational Compilations}
\label{sec:data}

It is well known that the dark energy domination epoch began at a late cosmic period. Then, the suitable data to study this dark component is at low redshifts. In this section, we explain the method to obtain the cosmological parameters $\Theta=\{h, \Omega_m, w_0,w_1\}$, where $h = H_0/(100 \, \text{km} \, \text{s}^{-1} \, \text{Mpc}^{-1})$ is the adimensional Hubble constant. We perform these analyses by considering the  $\Lambda$CDM model and the three bidimensional dark energy parameterisations described in Sec.~\ref{sec:DEbackground}. We begin by using the SNeIa Pantheon \cite{scolnic2018complete} and Cosmic Chronometers \cite{10.1093/mnras/sty260} compilations. Once this is done, we will implement our analyses with GW catalogues data \cite{abbott2019gwtc,Abbott:2020niy} to improve the constrains over the cosmological parameters. From the Bayes Theorem, considering some data vector $Y$ along with their uncertainties $\sigma_Y$ and the redshift associated with the data
\begin{equation}
\label{eqn:posterior}
\mathcal{P}(\Theta|Y,\sigma_Y,z) = \frac{\mathcal{P}(\Theta)\mathcal{P}(Y|\sigma_Y,\Theta,z)}{\mathcal{E}},
\end{equation}
where $\mathcal{P}(\Theta|Y,\sigma_Y,z)$ is the posterior probability, $\mathcal{P}(\Theta)$ the prior probability, $\mathcal{P}(Y|\sigma_Y,\Theta,z)$ the likelihood and
\begin{equation}
\mathcal{E} = \int d\Theta \mathcal{P}(\Theta) \mathcal{P}(Y|\sigma_Y,\Theta,z),
\end{equation}
is the evidence. The evidence is a normalisation constant and therefore it is not required to compute the best fit values of the parameter vector given some data. In order to compute the posterior probability, we need to know the likelihood. This function is given by \cite{hogg2010data}
\begin{equation}
\mathcal{P}(Y|\sigma_Y,\Theta,z) = \displaystyle\prod_{n=1}^N \frac{1}{\sqrt{2\pi \sigma_{Y_n}^2}} \exp \left( - \frac{1}{2} \left[ \frac{Y_n - Y_{n \, \text{theoretical}}}{\sigma^2_{Y_n}} \right]^2 \right),
\end{equation}
where $N$ is the number of samples, $Y_n$ is the $n$-th value of the data, $\sigma_{Y_n}$ its uncertainty, and $Y_{n \, \text{theoretical}}$ the theoretical value of $Y_n$ given some cosmological model.

If we apply the natural logarithm function to the posterior probability (\ref{eqn:posterior}), we will obtain the $\chi^2$ function from the likelihood
\begin{equation}
\chi^2 = \sum_{n=1}^N \left[ \frac{Y_n - Y_{n \, \text{theoretical}}}{\sigma^2_{Y_n}} \right]^2.
\end{equation}
By applying the natural logarithm, we will also have the factor
\begin{equation}
-\frac{1}{2} \sum_{n=1}^N \ln (2\pi \sigma_{Y_n}^2),
\end{equation}
which is constant. Therefore, the natural logarithm of the posterior is proportional to
\begin{equation}
\ln \mathcal{P}(\Theta|Y,\sigma_Y,z) \propto \ln \mathcal{P}(\Theta) - \frac{1}{2} \chi^2.
\end{equation}
If we want to determine the best fits and uncertainties of the parameter vector $\Theta$, we just need to know the prior probability and the $\chi^2$. The former function is assumed given some physical possible values for the parameters and the latter depends on the observational catalogue. In all cases, we will consider the prior probability given in Table \ref{table:priorProbability}.

Finally, when considering the combination of two or more catalogues, since each one is independent of the others, the natural logarithm of the posterior probability of the combination is the sum of the natural logarithms of each posterior. We can express this fact as
\begin{equation}
\ln \mathcal{P}(\Theta|Y,\sigma_Y,z) = \sum_{i}^M \ln \mathcal{P}(\Theta|Y_i,\sigma_{Y_i},z_i),
\end{equation}	
where $Y,\sigma_Y,z$ is data of the combination of the catalogues, and $Y_i,\sigma_{Y_i},z_i$ the data of the i-th catalogue.
\begin{table}[h]
	\centering
	\begin{tabular}{|c|c|}
		\hline
		Parameter & Prior probability distribution \\ \hline
		$h$  &  $\mathcal{U}(0.5,1.0)$ \\
		$\Omega_M$ & $\mathcal{U}(0.0,0.5)$ \\
		$w_0$ & $\mathcal{U}(-5.0,5.0)$ \\
		$w_1$ & $\mathcal{U}(-5.0,5.0)$ \\
		\hline
	\end{tabular}
	\caption{Prior probability distributions considered for the cosmological parameters. We will consider uniform probability distributions not to favour any value in the range.}
	\label{table:priorProbability}
\end{table}

\subsection{Pantheon supernovae catalogue}

The Pantheon catalogue \cite{scolnic2018complete} includes $N_{\text{Pantheon}}=1048$ events. These events are treated in an interval $z\in[0.010,
2.26]$. The catalogue gives the observational value of the distance modulus \cite{scolnic2018complete} with
\begin{equation}
	\mu_{\text{obs}} = m'_B - M,
\end{equation}
where $m'_B$ is the apparent magnitude in the B band given by \cite{scolnic2018complete,tripp1997two}
\begin{equation}
	m'_B = m_B + \alpha x_1 - \beta c + \Delta_M + \Delta_B,
\end{equation}
where $\Delta_M$ is a correction in distance based on the mass of the host-galaxy of the supernovae and $\Delta_B$ a distance correction from biases predicted from simulations \cite{scolnic2018complete}, and $\alpha$ and $\beta$ are coefficients of the relation between luminosity, and stretch and colour, respectively \cite{scolnic2018complete}. Therefore, $m'_B$ is a corrected value taking into account the systematic effects. 

Since the absolute magnitude $M$ and $H_0$ are degenerate parameters, we calibrated this catalogue with a value of $M$ in such a way that $H_0$ is the value given in \cite{Riess:2019cxk}. Because of this calibration, the best fits of $H_0$ in all the MCMCs that include the Pantheon catalogue will approach the prior value given in \cite{Riess:2019cxk}. However, this calibration does not give a preference for a certain cosmological model, since this value of $H_0$ is a model independent one \cite{Riess:2019cxk}. The theoretical distance modulus is given by
\begin{equation}
\label{eqn:distanceModulus}
\mu(z) = 5 \log_{10} \left( \dfrac{D_L(z)}{10 \, \text{pc}} \right),
\end{equation}
where $D_L(z)$ is is given by (\ref{eqn:luminosityDistanceComoving}). The best fit parameters $\Theta = \{ h, \Omega_M, w_0, w_1 \}$ can be calculated by maximizing the logarithm of the likelihood function \cite{hogg2010data}
\begin{eqnarray}
\label{eqn:maximumLikelihood}
\ln p(\mu(z)|z,\sigma,\Theta) = - \frac{1}{2} \left( \chi^2  + \sum_{n=1}^{N_{\text{Pantheon}}} \ln (2\pi \sigma_{n \, \text{obs}}^2 ) \right),
\end{eqnarray}
where the $\chi^2$ function is given by
\begin{equation}
\label{eqn:chiSquaredSNeIa}
\chi_{\text{Pantheon}}^2 = \sum_{n=1}^{N_{\text{Pantheon}}} \left(\frac{\mu(z_n,\Theta)-\mu_{\text{obs}}(z_n)}{\sigma_{\text{obs}}} \right)^2,
\end{equation}
with $\sigma_{n \, \text{obs}}^2$ the measured variances. We will refer to this catalogue as Pantheon in what it follows.

\subsection{Cosmic Chronometers Catalogue}

This database includes $N_{\text{CC}} = 51$ events in the redshift interval $z \in (0,2.36]$ \cite{10.1093/mnras/sty260}. Along with these data, we took the $H_0$ value from  \cite{Riess:2019cxk}, which is a model independent value. The other data points in this catalogue come from the differential age or clustering methods.

In the differential ages method, the age difference $\Delta t$, between two passively-evolving galaxies that born at the same time but separated by a redshift interval $\Delta z$, is measured \cite{jimenez2002constraining}. If we use the definition of the Hubble parameter, we can write $H(z) = -\dfrac{1}{1+z} \dfrac{dz}{dt}$. Therefore, we can estimate the Hubble factor with the differential ages method if we make the approximation $dz/dt \approx \Delta z/\Delta t$.

On the other hand, data points from the clustering method come from baryon acoustic oscillations (BAO) measurements. Some of these points are biased because their Hubble factor is estimated using the sound horizon $r_d$ (the maximum comoving distance at which sound waves can travel at redshift $z_d$) at redshift $z_d$, which depends on the cosmological model \cite{10.1093/mnras/sty260}. Also, we have 20 events measured using this method  \cite{10.1093/mnras/sty260}. So, they may introduce a bias on the results of the best fits and uncertainties of the cosmological parameter vector. This bias will occur in the analyses including the cosmic chronometers catalogue.

In this case, since this catalogue gives measurements of the Hubble parameter $H(z)$, the $\chi^2$ function is given by
\begin{equation}
\label{eqn:chiSquaredHubbleParameter}
\chi_{\text{CC}}^2 = \sum_{n=1}^{N_{\text{Hubble}}} \left(\frac{H(z_n,\Theta)-H_{\text{obs}}(z_n)}{\sigma_{n \, \text{obs}}} \right)^2,
\end{equation}
where and $H_{\text{obs}}$ and $\sigma_{n \, \text{obs}}$  correspond to the observations of the Hubble factor and their errors, respectively. From now on, we will refer to this catalogue as CC.


\subsection{Gravitacional Waves Catalogue}

Furthermore, in this work we consider data coming from GWs. This will be to update the previous contraints on the bidimensional dark energy models on their cosmological parameters \cite{Escamilla-Rivera:2016qwv}. In addition to this, we will see the constraining power of GW and how many standard siren events we need to have a determination of $H_0$ with a 1\% relative error. We will do the former with real GW data coming from the GWTC-1 \cite{abbott2019gwtc} and GWTC-1 \cite{Abbott:2020niy} catalogues of the LIGO-Virgo collaboration. And the latter will be done with simulated mock data taking the $\Lambda$CDM standard model. We will work with the median values reported in the LIGO-Virgo papers GWTC-1 \cite{abbott2019gwtc} and GWTC-2 \cite{Abbott:2020niy}, where they have already taken into account the GW selection effects.


\subsubsection{Gravitational-Wave Transient Catalogue GWTC-1}

This catalogue includes $N_{\text{GWTC-1}}=11$ confident detection events from gravitational-wave searches for coalescing compact binaries, all of them with masses $M > 1 M_{\odot}$. These events, observed by LIGO \footnote{\url{https://www.ligo.caltech.edu}} and Virgo\footnote{\url{http://www.virgo.infn.it}}, come from the first and second observing runs \cite{abbott2019gwtc}. In Table III from \cite{abbott2019gwtc}, the 11 events are reported along with their observational siren distances $D_S$ in Mpc, chirp masses $\mathcal{M}$ and other several observational variables, all of them with their corresponding $90 \%$ credible variances. 

As we saw in Sec. \ref{sec:cosmologyDarkSirens},  to compute the posterior probability with dark sirens, we need their chirp masses in the detector reference frame, its siren distances, and their signal to noise ratios. We consider the median values reported in Table III of \cite{abbott2019gwtc} in the case of the siren distances. LIGO-Virgo reported the intrinsic chirp masses computed by dividing the detector frame chirp masses by $1+z$, with $z$ the redshift computed from the siren distance distribution and a flat $\Lambda$CDM model \cite{abbott2019gwtc}. Hence, in order to recover the detector frame chirp masses, we multiplied the median values of the chirp masses reported in table III of \cite{abbott2019gwtc} by $1+z$, with $z$ the median values reported in the same Table. Finally, we consider the signal to noise ratios reported in Table I from \cite{abbott2019gwtc}.


\subsubsection{Gravitational-Wave Transient Catalogue GWTC-2}

This second GW catalogue includes $N_{\text{GWTC-2}}=39$ candidate GW events, with less than $10\%$ of contamination fraction from the first half of the third observable run \cite{Abbott:2020niy}. In Table VI from \cite{Abbott:2020niy}, these events measured by LIGO-Virgo are presented with their observational siren distances $D_S$ in Gpc\footnote{We need to perform a units conversion to be consistent with the GWTC-1 database, where the $D_S$ is reported in Mpc.}, their signal to noise ratios $\rho$, their chirp masses in the reference frame of the detector $\mathcal{M}$, along with other several observational variables, all of them with their corresponding $90\%$ credible variances. 

In this case, we consider the median values reported in Table VI from \cite{Abbott:2020niy} in the case of the siren distances. Following the procedure of the data gathered from the GWTC-1, in order to recover the detector frame chirp masses, we multiplied the median values of the chirp masses reported in Table VI of \cite{Abbott:2020niy} by $1+z$ with $z$ the median valued reported in the same Table. Finally, we consider the signal to noise ratios reported in Table IV from \cite{Abbott:2020niy}.


\subsubsection{GW data considered}

We will consider the following events from GWCT-1 \cite{abbott2019gwtc} and GWCT-2 \cite{Abbott:2020niy}:
\begin{itemize}
	\item The standard siren GW170817.
	\item The candidate standard siren GW190521.
	\item Dark sirens coming from BBH mergers, except GW190521 which is a standard siren candidate.
\end{itemize}

GWTC-1 \cite{abbott2019gwtc} and GWTC-2 \cite{Abbott:2020niy} have two confirmed or standard siren candidates: GW170817 and GW190521. The former has an associated electromagnetic counterpart given by the gamma ray burst GRB-170817A \cite{abbott2017gravitational} which has a redshift $z=0.0099$ \cite{abbott2019properties}. The latter has an electromagnetic candidate counterpart given by  three-dimensional spatial location of \textit{ZTF19abanrhr}, which has a redshift $z=0.438$ \cite{chen2020standard}. By taking their siren distances from the GWTC-1 \cite{abbott2019gwtc} and GWTC-2 \cite{Abbott:2020niy} catalogues, we can write the $\chi^2$ of these events by
\begin{equation}
	\chi^2_{\text{standard sirens}} = \sum_{n=1}^2 \left(\frac{D_S(z_n,\Theta)-D_{S \, \text{obs}}(z_n)}{\sigma_{n \, \text{obs}}}\right)^2,
\end{equation}
where $D_{S \, \text{obs}}$ and $\sigma_{n \, \text{obs}}$ denote the siren distances and their errors, respectively and taken from Table III in \cite{abbott2019gwtc} and Table IV from \cite{Abbott:2020niy}.

In addition to these events, we worked with the dark sirens produced in BBH mergers, which are 46 of the 50 events from GWTC-1 \cite{abbott2019gwtc} and GWTC-2 \cite{Abbott:2020niy}. However, since GW190521 is a candidate standard siren, we will not include it in the dark siren analyses.

We do not have a $\chi^2$ function for the case of dark sirens since they do not have an independent measurement of their redshift. However, we can work with the posterior probabilities discussed in Sec.\ref{sec:cosmologyDarkSirens} and combine these data with the standard sirens by adding their natural logarithms of their individual posteriors as we discussed in the beginning of this section.

We will refer to the data described in this subsection as GW and it includes $N_{\text{GW}}=47$ events. Since GWTC-2 \cite{Abbott2020prospects} reports events with less than $10\%$ of contamination fraction \cite{Abbott:2020niy}, the best fit values and their confidence regions will be biased a little. However, this bias will be minimal due the density of data when combining it with the Pantheon and CC catalogues, because the GWTC-2 that contains the bias represents just the $3.39\%$ of the total Pantheon+CC+GW database. 


\subsubsection{GW mock data generated}

In addition to the analyses performed with real GW data from GWTC-1 \cite{abbott2019gwtc} and GWTC-2 \cite{Abbott:2020niy}, we simulated GW standard sirens to see how many of those events we would need to constraint the Hubble constant $H_0$ with a relative error within 1\%, i.e. to constraint this value with a relative error similar to the one reported in \cite{Riess:2019cxk}. In order to generate the mock catalogue, we need to assume a cosmological model which will be the standard $\Lambda$CDM model. We also need to assume some values for $h$ and $\Omega_M$ and we will consider the best fits given by $\Lambda$CDM and the Pantheon catalogue reported in Table \ref{table:bestFitValues} ($h = 0.7284$ and $\Omega_M = 0.285$). We consider these results, since the Pantheon catalogue was calibrated with the $H_0$ from \cite{Riess:2019cxk} and therefore these cosmological parameters follow a cosmology given by this value of $H_0$. We will generate the mock catalogue in the redshift range $z \in (0,2.3)$ in comparison the same range used in \cite{Riess:2019cxk}. With these assumptions we can simulate the siren distances
\begin{equation}
	\label{eqn:mockSirenDistances}
	D_{S \, \text{simulated}}(z) = D_{S \, \Lambda\text{CDM}}(z) + \mathcal{N}(0,\sigma),
\end{equation}
where $D_{S \, \Lambda\text{CDM}}$ is the siren distance assuming $\Lambda$CDM and the cosmological parameters previously discussed and $\mathcal{N}(0,\sigma)$ the Normal Gaussian Probability distribution with mean 0 and standard deviation $\sigma$. These standard deviations will be given by the simulated errors in the siren distance which will be generated with a uniform probability distribution in order to have relative errors between 7\% and 30\%. Therefore, if future standard sirens have a smaller error, we will need less of them to get a constrain of $H_0$ with a relative error of 1\% than the value given in this paper.

Finally, the mock redshifts will be generated with the normalised intrinsic merger rate $\dot{n}(z)$ as probability distribution. We will consider the total intrinsic merger rate by adding the one corresponding to BBH, BNS, and BHNS. Afterwards we will simulate their siren distances using (\ref{eqn:mockSirenDistances}).


\section{Methodology}
\label{sec:results}

We performed a Markov-chain Monte Carlo (MCMC) analysis to compute the posteriors using a modified version of the \texttt{emcee} code\footnote{\url{https://emcee.readthedocs.io}} \cite{foreman2013emcee} interfaced to the publicly available sampling code \texttt{MontePython-v3}\footnote{\url{https://github.com/brinckmann/montepython\_public}} 
\cite{Audren:2012wb,Brinckmann:2018cvx}. 
We obtained plots of the 1-$\sigma$ and 2-$\sigma$ confidence regions and best fit values with their 1-$\sigma$ uncertainties using the \texttt{GetDist} package\footnote{\href{https://getdist.readthedocs.io/}{https://getdist.readthedocs.io/en/latest}} \cite{Lewis:2019xzd}. 
We sampled over the parameters $\{h, \Omega_m, \Theta, \Xi_0, n\}$ using the Metropolis-Hastings method and with a Gelman-Rubin \cite{Gelman:1992zz} convergence criteria around $R-1<0.03$.

\subsection{Statistical analysis}

The best fit values and the 1-$\sigma$ uncertainties for the $h$, $\Omega_m$, $w_0$, and $w_1$ parameters using the Pantheon and CC databases and their combination are reported in Table \ref{table:bestFitValues}. The best fit values and the 1-$\sigma$ uncertainties for the $h$, $\Omega_m$, $w_0$, $w_1$ parameters for the case of the GW database along with their combination with Pantheon and CC are reported in Table \ref{table:bestFitValuesGW}. Finally, we generated $N=50,200,1000,2000$ mock siren events. The best fit values and the 1-$sigma$ uncertainties for the $h$ and $\Omega_m$ parameters for $\Lambda$CDM and this mock catalogue are reported in Table \ref{table:mockSiren}.

\begin{figure*}
	\includegraphics[width=7.6cm]{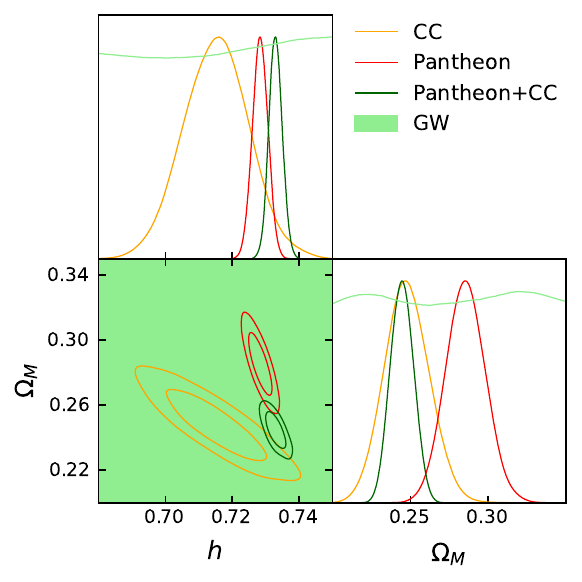}
	\includegraphics[width=7.6cm]{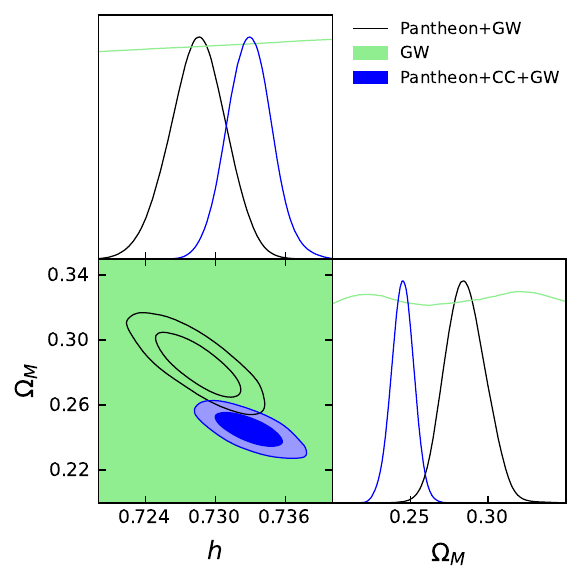}
	\caption{1-$\sigma$ and 2-$\sigma$ confidence contours for the cosmological parameters $h$ and $\Omega_m$ for the $\Lambda$CDM model. \textit{Left:} Analyses performed using Pantheon (red colour), CC (orange colour), Pantheon+CC (green colour), and GW (light green colour) catalogues. \textit{Right:} Analyses performed using GW (light green colour), Pantheon+GW (black colour),  and Pantheon+CC+GW (blue colour) catalogues.
	}
	\label{fig:LCDM-Overlapping-Pantheon+CC}
\end{figure*}

\begin{figure*}
	\includegraphics[width=7.6cm]{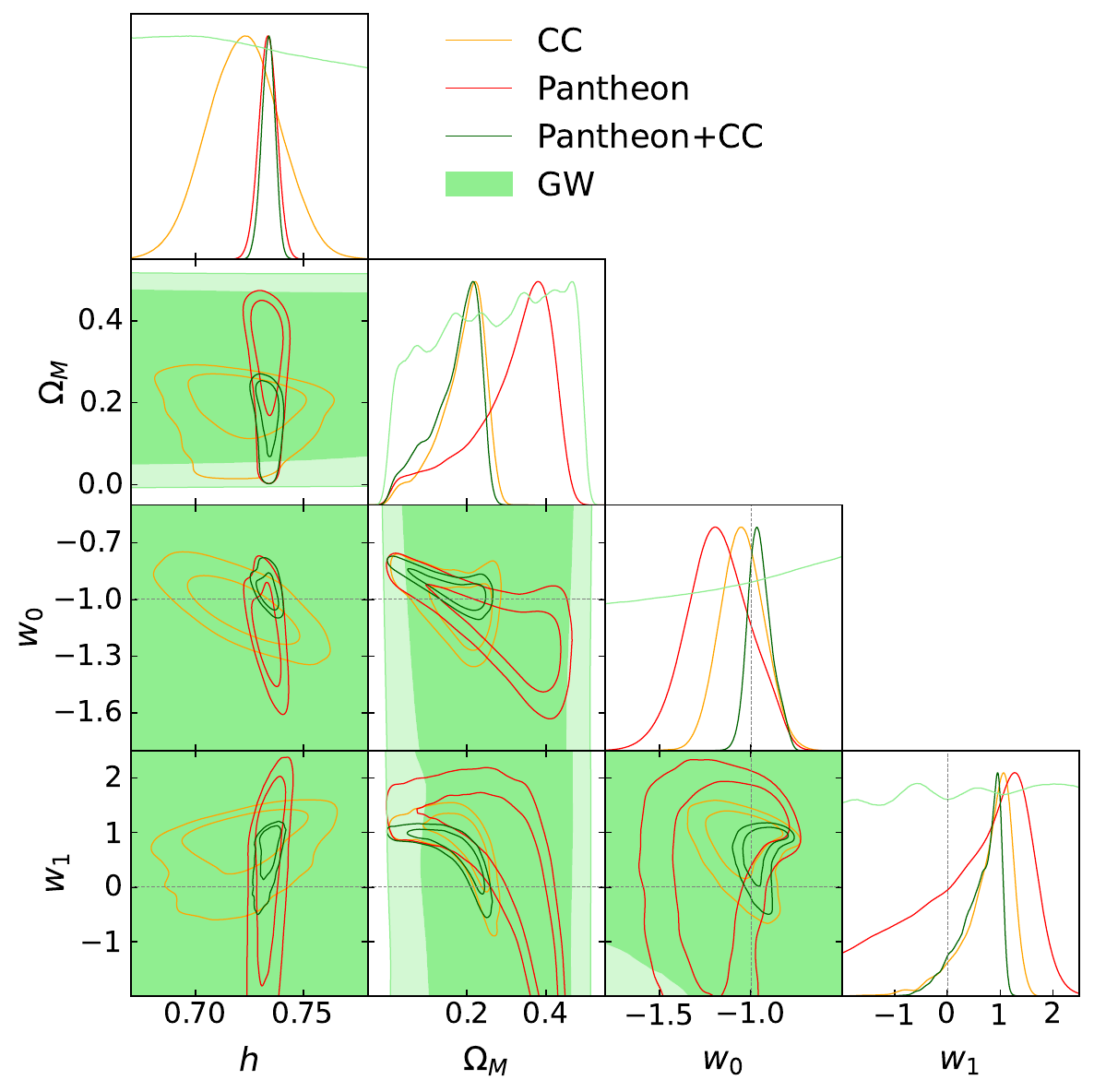}
	\includegraphics[width=7.6cm]{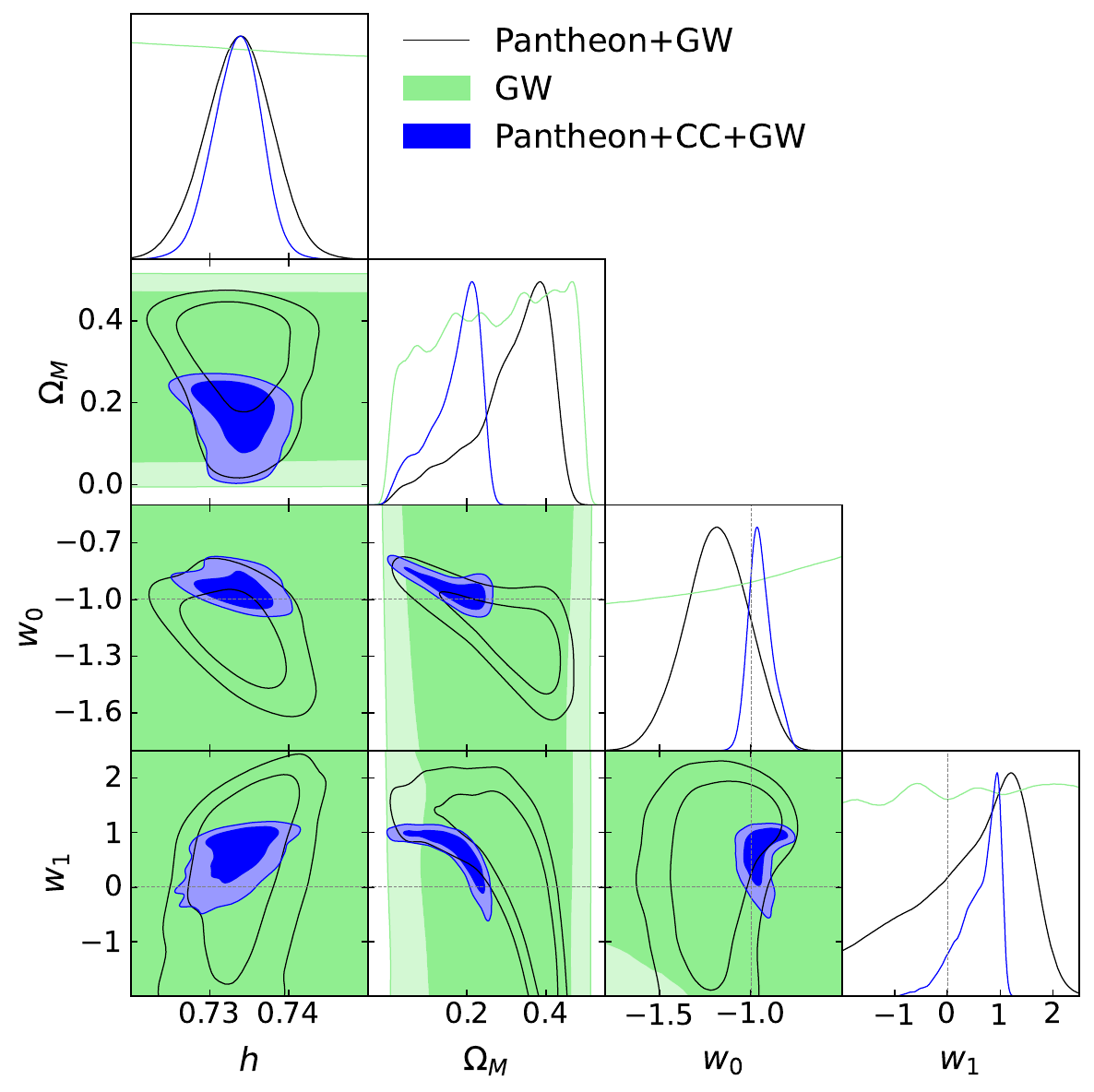}
	\caption{1-$\sigma$ and 2-$\sigma$ confidence contours for the cosmological parameters $h$, $\Omega_m$,  $w_0$ and $w_1$ for the CPL model. \textit{Left:} Analyses performed using Pantheon (red colour), CC (orange colour), Pantheon+CC (green colour), and GW (light green colour) catalogues. \textit{Right:} Analyses performed using GW (light green colour), Pantheon+GW (black colour),  and Pantheon+CC+GW (blue colour) catalogues.
	}
	\label{fig:CPL-Overlapping-Pantheon+CC}
\end{figure*}

\begin{figure*}
	\includegraphics[width=7.6cm]{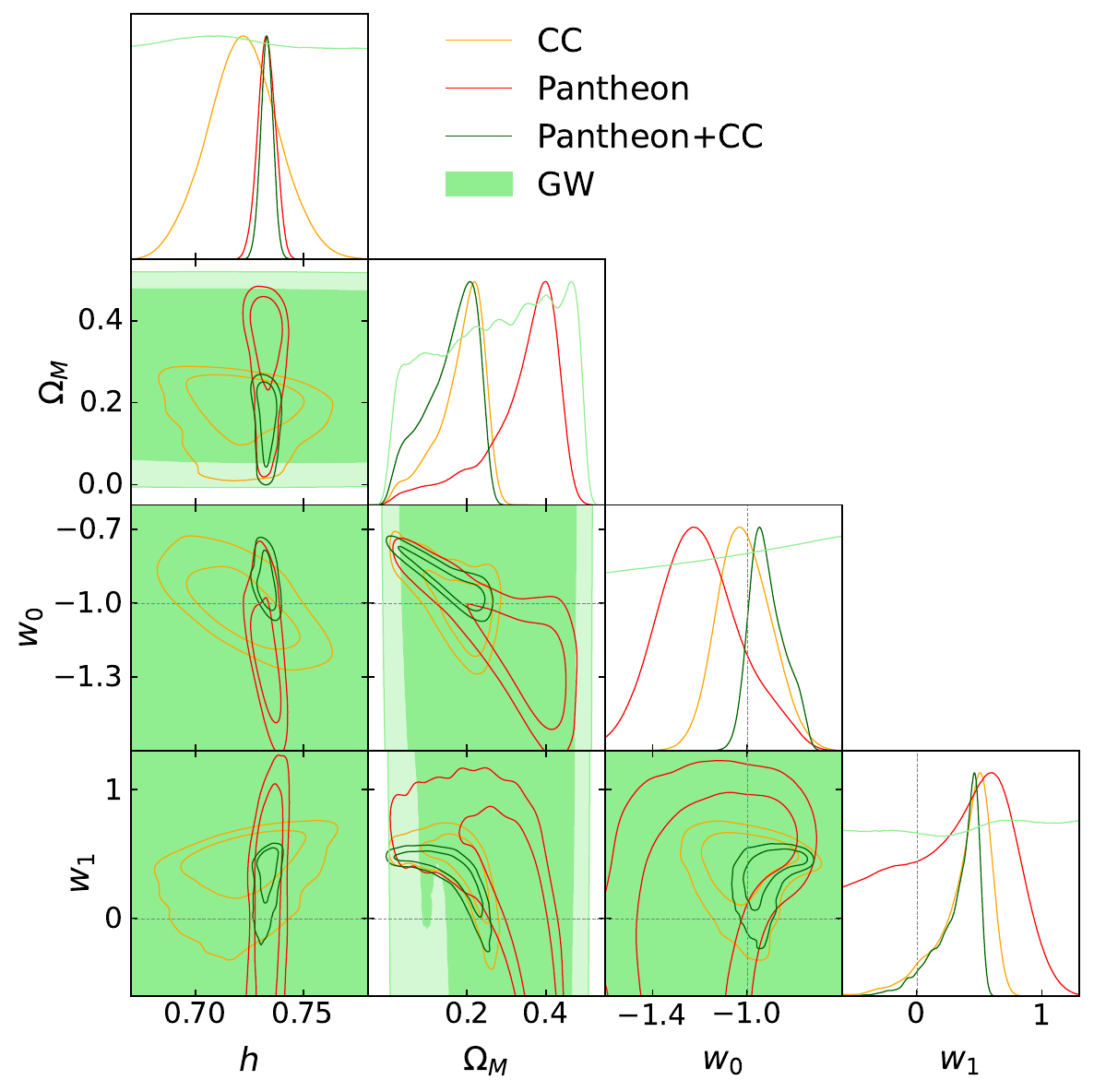}
	\includegraphics[width=7.6cm]{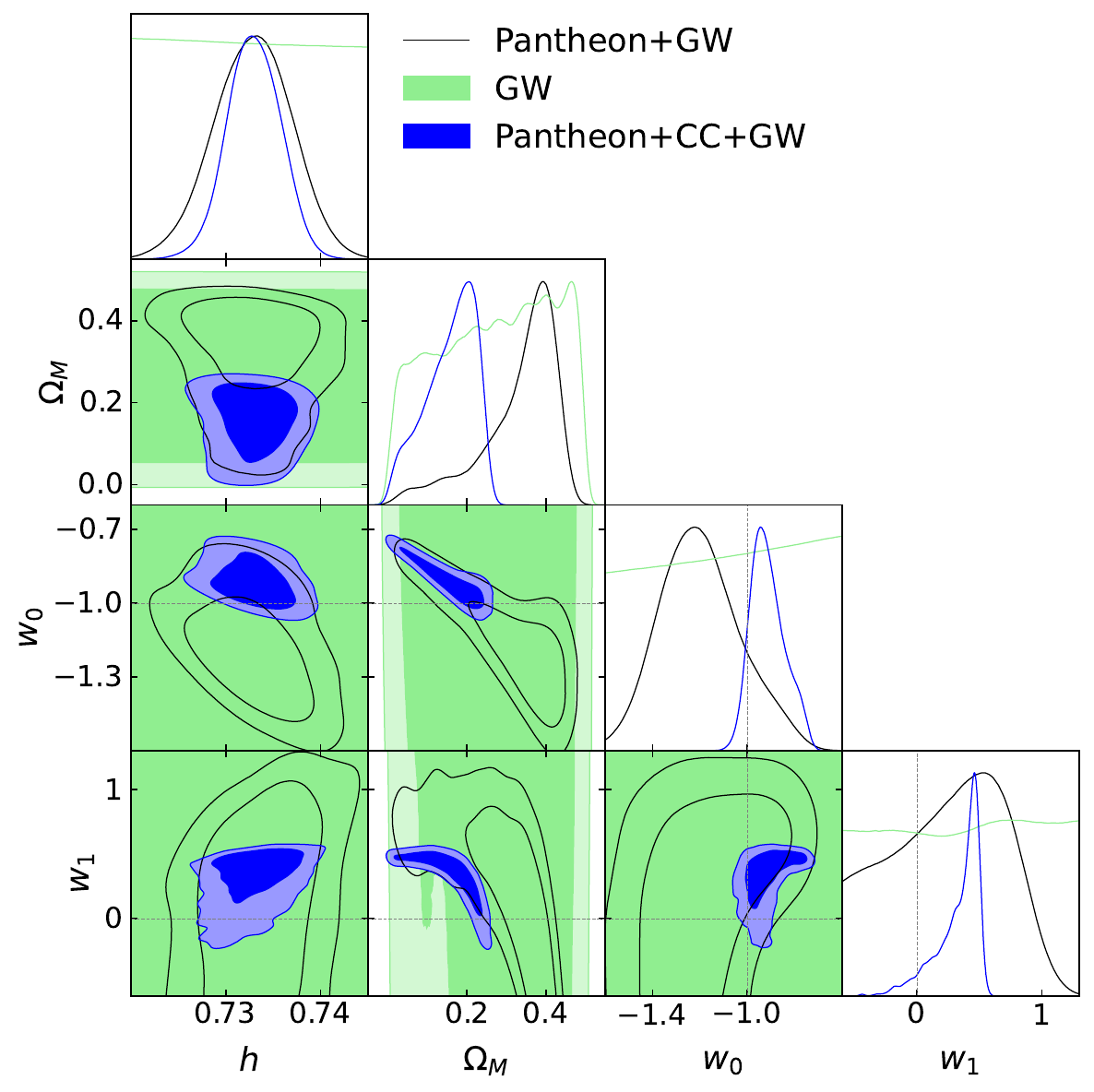}
	\caption{1-$\sigma$ and 2-$\sigma$ confidence contours for the cosmological parameters $h$, $\Omega_m$,  $w_0$ and $w_1$ for the BA model. \textit{Left:} Analyses performed using Pantheon (red colour), CC (orange colour), Pantheon+CC (green colour), and GW (light green colour) catalogues. \textit{Right:} Analyses performed using GW (light green colour), Pantheon+GW (black colour),  and Pantheon+CC+GW (blue colour) catalogues. 
	}
	\label{fig:BA-Overlapping-Pantheon+CC}
\end{figure*}

\begin{figure*}
	\includegraphics[width=7.6cm]{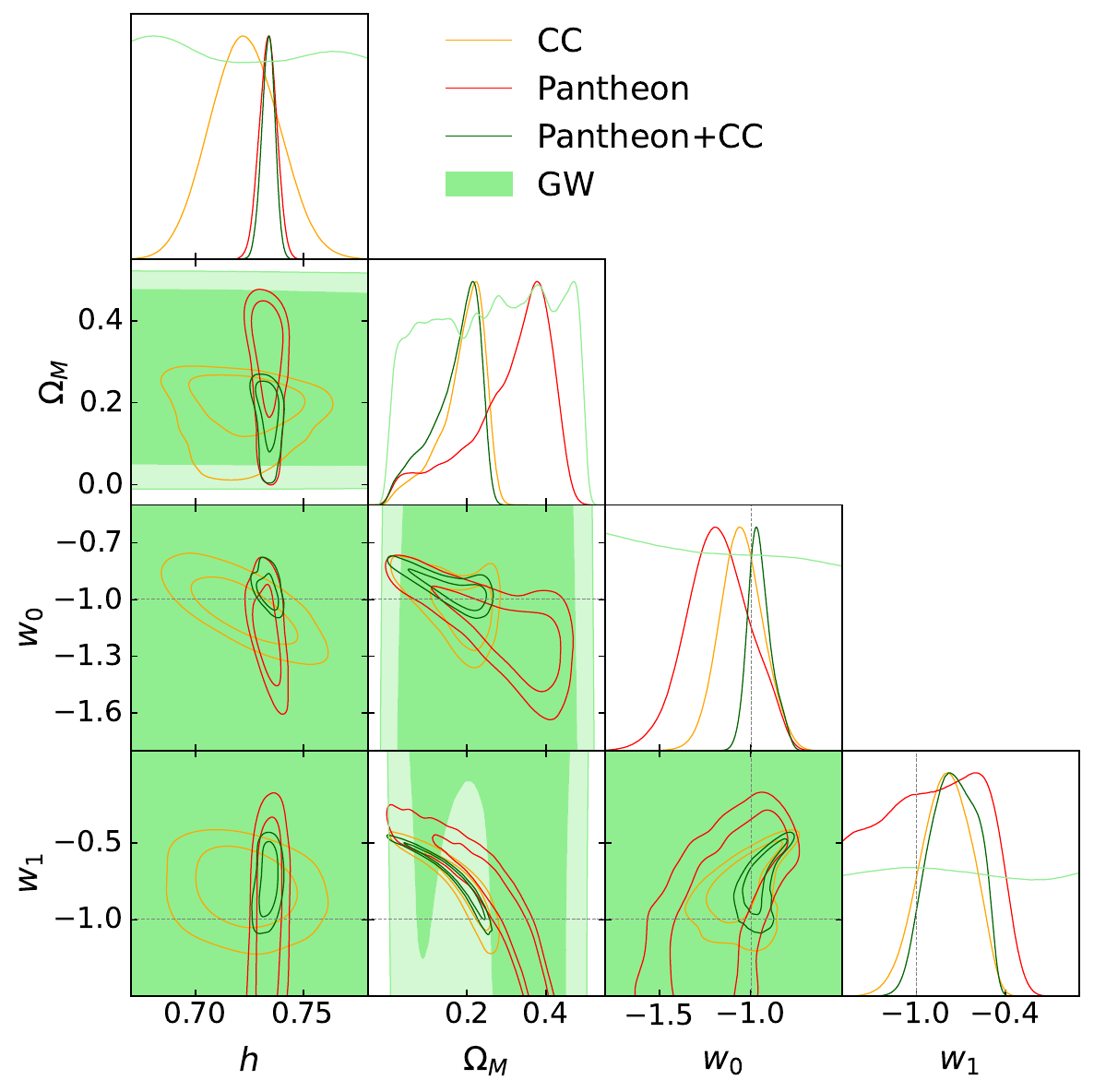}
	\includegraphics[width=7.6cm]{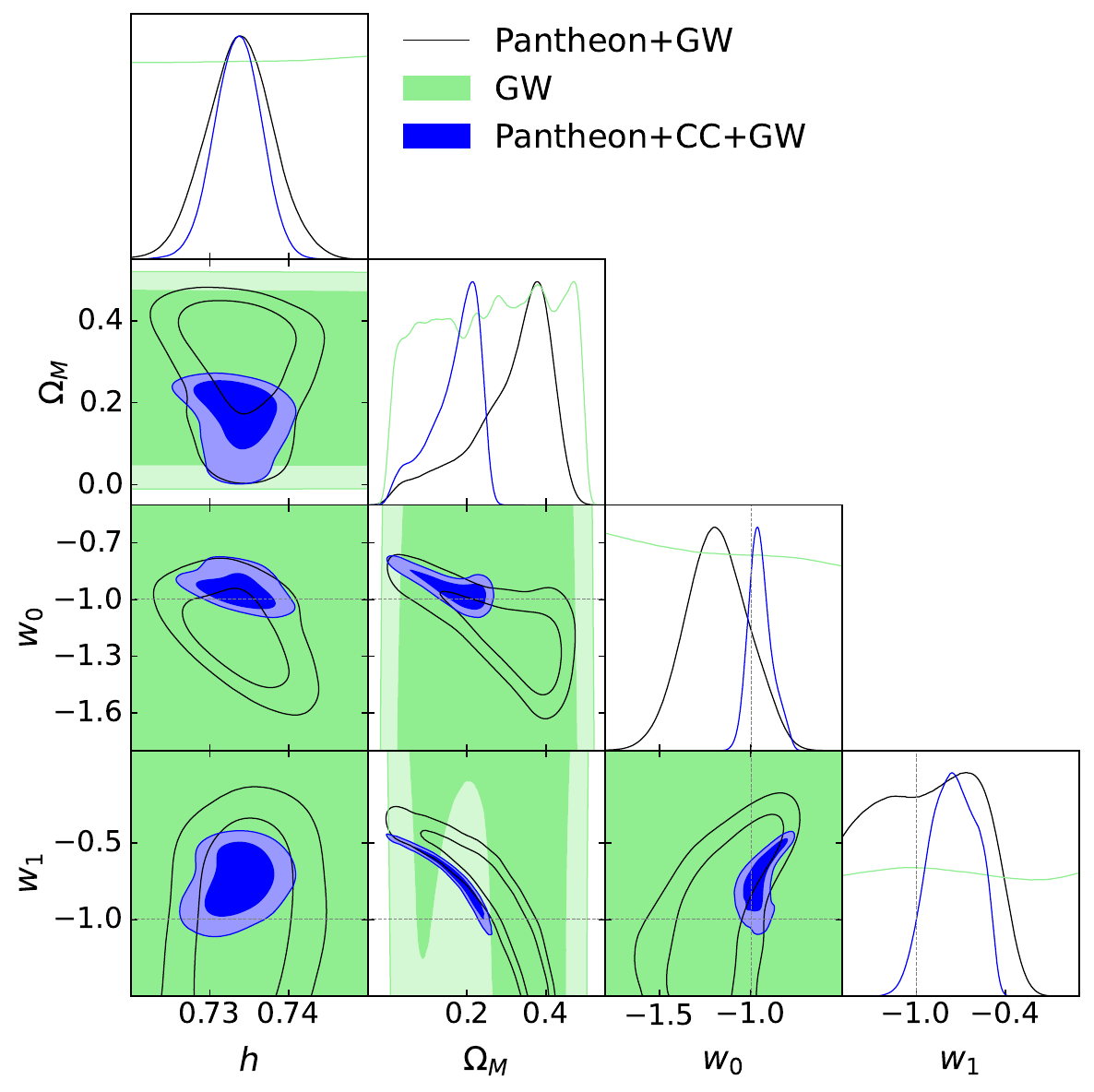}
	\caption{1-$\sigma$ and 2-$\sigma$ confidence contours for the cosmological parameters $h$, $\Omega_m$,  $w_0$ and $w_1$ for the LC model. \textit{Left:} Analyses performed using Pantheon (red colour), CC (orange colour), Pantheon+CC (green colour), and GW (light green colour) catalogues. \textit{Right:} Analyses performed using GW (light green colour), Pantheon+GW (black colour),  and Pantheon+CC+GW (blue colour) catalogues. We consider $w_1=w_c$ for this case.
	}
	\label{fig:LC-Overlapping-Pantheon+CC}
\end{figure*}

\begin{figure*}
	\centering
	\includegraphics[width=10.0cm]{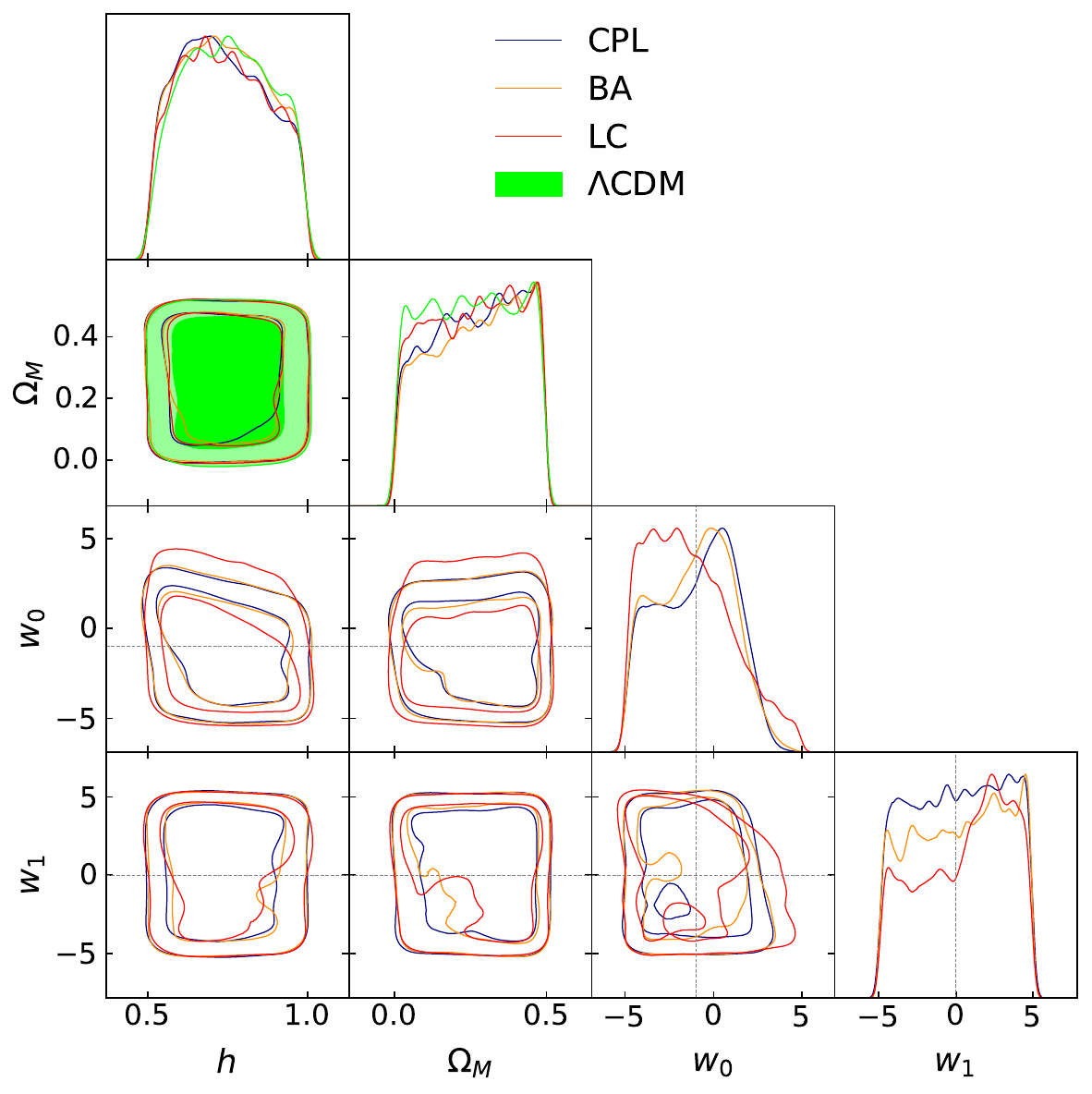}
	\caption{1-$\sigma$ and 2-$\sigma$ confidence contours for the cosmological parameters $h$, $\Omega_M$, $w_0$, $w_1$ ($w_1=w_c$ for the LC model) for the four models and the GW full catalogue. }
	\label{fig:overlap_deltas/models-Overlapping-GW}
\end{figure*}

\begin{figure*}[h]
	\centering
	\includegraphics[width=9.0cm]{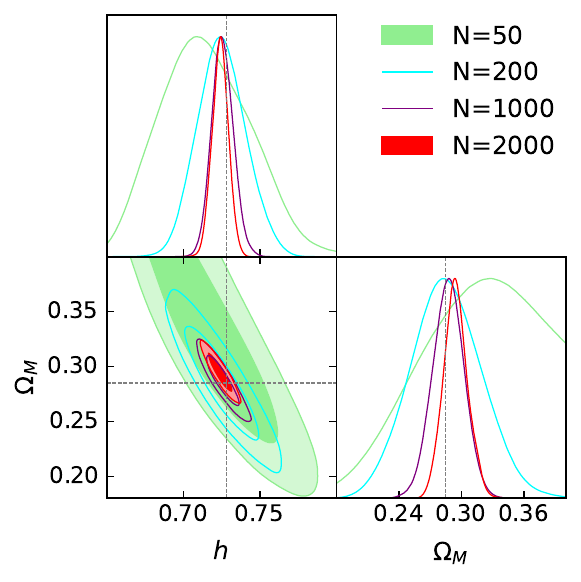}
	\caption{1-$\sigma$ and 2-$\sigma$ confidence contours for the cosmological parameters $\Theta = \{ h, \Omega_M \}$ for $\Lambda$CDM and the mock standard sirens. $N$ refers to the number of mock standard sirens.
	}
	\label{fig:mock-siren}
\end{figure*}

We also included the plots at 1-$\sigma$ and 2-$\sigma$ confidence regions for the four models  $\Lambda$CDM model and the three bidimensional dark energy parametrisations). We presented two plots per model: (i) with the analyses from Pantheon, CC, their combination Pantheon+CC and GW and (ii) with the analyses from GW and their combinations with Pantheon and CC (Pantheon+GW and Pantheon+CC+GW). The plots are arranged as: Figure \ref{fig:LCDM-Overlapping-Pantheon+CC}  for the standard $\Lambda$CDM model, Figure \ref{fig:CPL-Overlapping-Pantheon+CC} for CPL, Figure \ref{fig:BA-Overlapping-Pantheon+CC} for BA, and Figure \ref{fig:LC-Overlapping-Pantheon+CC} for LC models. Then, we included a plot with the four models and the GW catalogue. This plot is given in figure \ref{fig:overlap_deltas/models-Overlapping-GW}. Finally, we plot the confidence regions for the case of the GW mock catalogue. This plot is given in Figure \ref{fig:mock-siren}.

\subsection{Bayesian Evidence Criterion Analysis}

To quantify to what extent the improvement in our fit for the full Pantheon+CC+GW catalogue described warrants the increase in the model complexity compared to the $\Lambda$CDM model, we compute the Bayes factor --the ratio of the evidences for the baseline model ${\cal M}_{\rm L}$ with respect to the extended ${\cal M}_{\rm E}$-- as \cite{Jeffreys:1939xee}: 
\begin{eqnarray}
B_{EL} \equiv \frac{\int d\boldsymbol{\theta}_L\, \pi(\boldsymbol{\theta}_L \vert {\cal M}_L) {\cal L}(\mathbf{x} \vert \boldsymbol{\theta}_L,{\cal M}_L)\,}{\int d\boldsymbol{\theta}_E\, \pi(\boldsymbol{\theta}_E \vert {\cal M}_E) {\cal L}(\mathbf{x} \vert \boldsymbol{\theta}_E,{\cal M}_E)\,}\,
\label{eq:bayesfactor} = \frac{\mathcal{E}(\mathcal{M}_L)}{\mathcal{E}(\mathcal{M}_E)},
\end{eqnarray}
where $\pi(\boldsymbol{\theta}_{\rm E,\,L})$ is the prior for the parameters ${\theta}_{\rm E,\,L}$ and ${\cal L}(\mathbf{x} \vert \boldsymbol{\theta}_{\rm E,\,L})$ the likelihood of the data given the model ${\cal M}_{\rm E,\,L}$. The extent to what the baseline model ${\cal M}_{\rm L}$ is preferred over the extended ${\cal M}_{\rm E}$ can be qualitatively assessed using the Jeffreys scale reported in Table~\ref{tab:kassraftery}. To compute the evidence, we use a nested sampling algorithm implemented in a modified version of the \texttt{dynesty} software ~\cite{feroz2009multinest,speagle2020dynesty,skilling2004nested,skilling2006nested}, which has been seen viable for cosmological applications ~\cite{liddle2006present,Escamilla-Rivera:2016qwv}.

\begingroup
\begin{center}
	\begin{table}
		\centering
		\begin{tabular}{|c||c|}
			\hline
			Information Criteria & Preference for model ${\cal M}_i$ \\ \hline \hline
			~~$0 < \ln B_{\text{E,L}} < 1 \qquad$ & Weak \\ \hline
			$1 \leq \ln B_{\text{E,L}} < 2.5$ & Definite \\ \hline
			$2.5 \leq \ln B_{\text{E,L}} < 5$ & Strong \\ \hline
			$5 \leq \ln B_{\text{E,L}} \qquad$ & Very strong \\
			\hline
		\end{tabular}
		\caption{Revised Jeffreys scale used to interpret the values of $\ln B$ obtained when comparing two competing models through their Bayesian evidence. A value of $\ln B>0$ indicates that the extended model (E) is favoured with respect to the $\Lambda$CDM baseline (L) model. }
		\label{tab:kassraftery}
	\end{table}
\end{center}
\endgroup

\subsection{Discussion}

As it can be seen in the plots, the $\Lambda$CDM model behaviour ($w_0 =-1$ and $w_1=0$ for all the parametrisations, except for the LC model $w_1=0$, $w_c=-1$) lies within the 2-$\sigma$ confidence contours for all cases. This shows that our results with all catalogues and combinations are consistent with the standard $\Lambda$CDM model.

According to the results in Table \ref{table:bestFitValuesGW}, the GW catalogue is capable of constraining the value of the cosmological parameters. However, they have big relative errors, between 20\% and 25\% for the $h$ parameter. If we remember that this catalogue has 2 standard sirens and 45 dark sirens, we can conclude that dark sirens do not have the same constraining power as CC since our CC catalogue with 52 events has relative errors between 1\% and 3\% for $h$. 

However, the results with the GW catalogue are consistent with the other catalogues and their combinations. Moreover, as we can see in Tables \ref{table:bestFitValues} and \ref{table:bestFitValuesGW}, results with Pantheon and CC did not change too much when adding the GW catalogue. This indicates that the GW catalogue did not have a strong statistical effect in the analyses. This is expected since Pantheon has 1048 events  in comparison to the 47 GW events. Additionally, the posterior probabilities, best fits and confidence regions computed with the nested sampling algorithm were very close (with differences of less than 1\%) to the ones from the MCMC algorithm. Hence, apart from computing the evidence, the nested sampling algorithm served as a validation of our results.

Nevertheless, future detections of GW, both standard and dark sirens, will help to constrain the $H_0$ constant and to study the Hubble constant tension problem. In Table \ref{table:mockSiren}, we can see that we need around 1000 of standard sirens to constrain $H_0$ with a relative error of 1\%. This is a really small number compared to the $10^5$ required dark sirens to have a similar performance \cite{ding2019cosmological}.

In Table \ref{table:InformationCriteria1}, we can see that the standard $\Lambda$CDM model is the preferred one. It shows a strong evidence against CPL and a very strong evidence against BA and LC. Therefore, by considering the full catalogue Pantheon+CC+GW, these alternate dark energy models are not able to challenge the standard $\Lambda$CDM. 

As we can notice, we need more GW data in order to study problems in cosmology. Even so, the future with GW cosmology is high promising. We can compute the cosmological parameters in an independent way in comparison to CC or SNeIa datasets. Hence, even though current GW data gives big relative errors in $h$ and the other parameters, they will be a key future tool with a much bigger number of standard and dark sirens coming from the Einstein telescope \cite{sathyaprakash2012scientific} and LISA \cite{babak2011fundamental} future third generation detectors.


\section{Conclusions}
\label{sec:conclusions}

In this paper we present an update to the analysis of three dark energy parameterisations using the new GW databases described. We considered current Pantheon+CC+GW data. Where the GW data comes from the GWTC-1 \cite{abbott2019gwtc} and GWTC-2 \cite{Abbott:2020niy} catalogues. With this new database, $\Lambda$CDM had a very strong preference against the BA and LC models and a strong preference against CPL model, with the Bayes evidence criterion.

Furthermore, we proposed a method to constrain the cosmological parameters with dark sirens given by Ding et al. \cite{ding2019cosmological}. This method does not require an independent measurement of the redshift of the samples. However, current data gave a relative error between 20\% and 25\% for the $h$ parameter which is much higher than the less than 3\% given with the CC catalogue.

This method along with the one with standard sirens will be fundamental in the study of cosmology with GWs once the Einstein Telescope \cite{sathyaprakash2012scientific} and LISA \cite{babak2011fundamental} detectors will give us a much higher number of dark and possibly standard sirens. We expect to need $10^5$ dark sirens \cite{ding2019cosmological} and 1000 standard sirens to constrain $h$ with a 1\% relative arror.

Additionally, from computing the Hubble constant $H_0$, the methods presented in this paper can be used in a wide variety of models derived from GR, this since modified and extended theories of gravity have a siren distance and luminosity distance inequality \cite{Mitra:2020vzq} as we can see in (\ref{eqn:sirenDistance}). GWs constitute an independent cosmological observable that will be able to quantify deviations from GW, therefore, these observations will help future efforts to see if these models can constitute an alternative to the $\Lambda$CDM model. 
These studies will be reported elsewhere.

\begin{table*}
	{\centering
		\begin{tabular}{|c|c|c|c|c|c|}
			\hline
			Model & Database & $h$ & $\Omega_M$ & $w_0$ & $w_1$ \\ \hline
			\multirow{3}{*}{$ \Lambda$CDM} & CC & $0.7152\pm 0.0099$ & $0.248\pm 0.014$ & - & - \\
			& Pantheon & $0.7284\pm 0.0023$ & $0.285\pm 0.013$ & - & - \\
			& Pantheon+CC & $0.7330\pm 0.0020$ & $0.2447\pm 0.0073$ & - & - \\ \hline
			\multirow{3}{*}{CPL} & CC & $0.722\pm 0.017$ & $0.186^{+0.071}_{-0.032}$ & $-1.05\pm 0.12$ & $0.77^{+0.54}_{-0.21}$ \\
			& Pantheon & $0.7335\pm 0.0042$ & $0.306^{+0.13}_{-0.044}$ & $-1.18\pm 0.17$ & $-0.05^{+1.9}_{-0.61}$ \\
			& Pantheon+CC & $0.7337\pm 0.0031$ & $0.168^{+0.078}_{-0.031}$ & $-0.950^{+0.053}_{-0.074}$ & $0.61^{+0.46}_{-0.17}$ \\ \hline
			\multirow{3}{*}{BA} & CC & $0.722 \pm 0.016$ & $0.184^{+0.073}_{-0.031}$ & $-1.01^{+0.10}_{-0.12}$ & $0.380^{+0.25}_{-0.090}$ \\
			& Pantheon & $0.7329\pm 0.0041$ & $0.336^{+0.11}_{-0.034}$ & $-1.20^{+0.15}_{-0.19}$ & $-0.40^{+1.3}_{-0.43}$ \\
			& Pantheon+CC & $0.7330\pm 0.0029$ & $0.159^{+0.085}_{-0.038}$ & $-0.915^{+0.057}_{-0.090}$ & $0.326^{+0.20}_{-0.061}$ \\ \hline
			\multirow{3}{*}{LC} & CC &$0.723\pm 0.016$ & $0.188^{+0.070}_{-0.028}$ & $-1.05\pm 0.12$ & $-0.80^{+0.19}_{-0.16}$ \\
			& Pantheon & $0.7336\pm 0.0041$ & $0.304^{+0.13}_{-0.047}$ & $-1.18\pm 0.17$ & $-1.18^{+0.75}_{-0.31}$ \\
			& Pantheon+CC & $0.7337^{+0.0032}_{-0.0029}$ & $0.170^{+0.078}_{-0.031}$ & $-0.952^{+0.050}_{-0.074}$ & $-0.75^{+0.19}_{-0.15}$ \\ \hline
		\end{tabular}
	}
	\caption{Best fit values for the four cosmological parameters with their one-$\sigma$ uncertainties. The best fit values are reported for the standard model $\Lambda$CDM and the three dynamical dark energy parametrisations The second column indicates the catalogue used.}
	\label{table:bestFitValues}
\end{table*}

\begin{table*}
	\centering
	\renewcommand{\arraystretch}{1.}
	\resizebox{\textwidth}{!}{%
		\begin{tabular}{|c|c|c|c|c|c|}
			\hline
			Model & Database & $h$ & $\Omega_M$ & $w_0$ & $w_1$ \\ \hline
			\multirow{3}{*}{$\Lambda$CDM} & GW & $0.75^{+0.13}_{-0.17}$ & $0.25\pm 0.14$ & - & -  \\
			& Pantheon+GW & $0.7285\pm 0.0024$ & $0.285^{+0.012}_{-0.014}$ & - & - \\
			& Pantheon+CC+GW & $0.7329\pm 0.0019$ & $0.2450\pm 0.0072$ & - & - \\ \hline
			\multirow{3}{*}{CPL} & GW & $0.74^{+0.12}_{-0.18}$ & $0.27^{+0.22}_{-0.12}$ & $-1.0^{+2.7}_{-3.0}$ & $0.2\pm 2.9$ \\
			& Pantheon+GW & $0.7336\pm 0.0042$ & $0.309^{+0.12}_{-0.044}$ & $-1.19\pm 0.17$ & $0.01^{+1.8}_{-0.67}$ \\
			& Pantheon+CC+GW & $0.7334^{+0.0032}_{-0.0028}$ & $0.169^{+0.079}_{-0.031}$ & $-0.948^{+0.052}_{-0.074}$ & $0.60^{+0.47}_{-0.18}$ \\ \hline
			\multirow{3}{*}{BA} & GW & $0.74^{+0.13}_{-0.17}$ & $0.271^{+0.22}_{-0.092}$ & $-1.1^{+2.5}_{-3.2}$ & $0.3\pm 2.9$  \\
			& Pantheon+GW & $0.7328\pm 0.0041$ & $0.333^{+0.11}_{-0.035}$ & $-1.20^{+0.15}_{-0.19}$ & $-0.33^{+1.2}_{-0.41}$ \\
			& Pantheon+CC+GW & $0.7330\pm 0.0028$ & $0.158^{+0.083}_{-0.040}$ & $-0.913^{+0.055}_{-0.088}$ & $0.327^{+0.20}_{-0.057}$ \\ \hline
			\multirow{3}{*}{LC} & GW & $0.74^{+0.12}_{-0.17}$ & $0.26^{+0.22}_{-0.15}$ & $-1.4^{+1.3}_{-3.3}$ & $0.6^{+4.2}_{-5.1}$    \\
			& Pantheon+GW & $0.7337\pm 0.0042$ & $0.309^{+0.13}_{-0.049}$ & $-1.19\pm 0.17$ & $-1.20^{+0.75}_{-0.32}$ \\
			& Pantheon+CC+GW & $0.7335\pm 0.0030$ & $0.170^{+0.077}_{-0.030}$ & $-0.949^{+0.049}_{-0.074}$ & $-0.75^{+0.18}_{-0.15}$ \\ \hline
	\end{tabular}}
	\caption{Best fit values for the four cosmological models discussed in Sec.\ref{sec:DEbackground} at 1-$\sigma$ uncertainties. In this case we consider the GW catalogues.}
	\label{table:bestFitValuesGW}
\end{table*}



\begin{table}
	\centering
	\begin{tabular}{|c|c|c|}
		\hline
		Model & 	$\ln \mathcal{E}$ &	$B_{\text{E L} }$  \\ \hline
		$\Lambda$CDM	& 336.01 &	- \\ \hline
		CPL	& 331.12 & 4.89 \\ \hline
		BA & 330.81 & 5.20 \\ \hline
		LC & 330.18 & 5.83 \\ \hline
	\end{tabular}
	\caption{Bayes evidence criterion for the three models dark energy against $\Lambda$CDM. $E$ denotes to the dark energy parametrisation, while $L$ is the baseline $\Lambda$CDM model.}
	\label{table:InformationCriteria1}	
\end{table}


\begin{table*}[h]
	\centering
	\begin{tabular}{|c|c|c|c|}
		\hline
		N & $h$ & $\Omega_M$ & Rel. err. $h$ \\ \hline
		50 & $0.715^{+0.027}_{-0.034}$ & $0.340^{+0.069}_{-0.079}$ & 4.7\% \\ \hline
		200 & $0.725\pm 0.016$ & $0.284^{+0.031}_{-0.037}$ & 2.2\% \\
		\hline
		1000 & $0.7254 \pm 0.0071$ & $0.288\pm 0.015$ & 0.98\% \\ \hline
		2000 & $0.7241\pm 0.0053$ & $0.294\pm 0.012$ & 0.73\% \\ \hline
	\end{tabular}
	\caption{Best fit values for the mock GW catalogue at 1-$\sigma$ uncertainties. To generate the mock catalogue, we used the $\Lambda$CDM catalogue. $N$ is the number of mock events and we also included the relative error in the constrain of $h$.}
	\label{table:mockSiren}
\end{table*}


\acknowledgments

CE-R is supported by DGAPA-PAPIIT UNAM Project TA100122 and acknowledges the Royal Astronomical Society as FRAS 10147.
This work is part of the Cosmostatistics National Group (\href{https://www.nucleares.unam.mx/CosmoNag/index.html}{CosmoNag}) project.
AN was supported by DGAPA-PAPIIT UNAM Project IA100220.


\bibliography{references}
\bibliographystyle{unsrt}


\appendix

\end{document}